\documentclass[english]{article}
\usepackage[T1]{fontenc}
\usepackage[latin9]{inputenc}
\usepackage{amsmath}

\makeatletter
\newcommand{\lyxaddress}[1]{
	\par {\raggedright #1
	\vspace{1.4em}
	\noindent\par}
}

\usepackage{babel}

\makeatother

\usepackage{babel}
\begin{document}
\title{Reflection Matrices for the $(n+1)(2n+1)$ Vertex Models: Diagonal
Solutions}
\author{A. Lima- Santos}
\maketitle

\lyxaddress{Universidade Federal de São Carlos, Departamento de Física, Caixa
Postal 676, CEP 13569-905, São Carlos, Brasil}
\begin{abstract}
We have find the diagonal $K$ matrix solutions of the reflection
equations for a class of vertex models. These models have $(n+1)(2n+1)$
vertices and are defined as two set of (n + 1) $R$ matrices, solutions
of the equations of Yang-Baxter equations. For a given value of $\text{n}$
we find n!-$\frac{1}{2}(n-2)(n-1)$ $K$ diagonal matrices. 

\numberwithin{equation}{section}
\end{abstract}

\section{Introduction }

In a previous work \cite{vieira0} we broke up the $A_{n-1}^{(1)}$
symmetry of the solution of the Yang-Baxter equations (YBE), the $R$-matrix
\cite{jimbo}. The symmetry break releases many parameters of the
model. These parameters are the values of the derivative of the entries
of the matrix $R(u)$, in a particular value of the parameter spectral
$\text{u}$, $\:u=0$ .

For $n>1$ the R (u) has $(n+1)(2n+1)$ entries,after setting some
parameters we can see that, $n(n+3)/2+3$ free parameters remain in
the two sets of n + 1 matrices $R$.

With so many parameters at our disposal, we can manipulate the solutions
until obtaining possible desired results. This freedom can be used
to open up many questions about integrality.

In this article after presenting the solutions of (YBE), with a difference
of two spectral parameters, we are going to treat the boundary Yang-Baxter
equations (BYBE ) or reflection equations (RE).

The RE are a natural extension of the YBE with an addition of one
element, the so-called reflection matrix or K-matrix. In the literature
on integrable models, we will find many works on this subject \cite{sklyanin,nepo,kulish}.
We will use a well-known approach to calculations \cite{dals,malara}
to find the diagonal solutions,. In the appendix A we also present
an essay of calculations for non-diagonal solutions. 

Solutions of (RE) are important in the study of critical phenomena
for the calculation of free energies and critical exponents \cite{batch}

\section{The Model}

The Yang-Baxter equation(YBE) can be seen as a matrix relation defined
in $End(V\otimes V\otimes V)$, where $V$ is an $n$- dimensional
complex vector space. In a particular and usual case, it reads: 
\begin{equation}
R_{12}(u)R_{13}(u+v)R_{23}(v)-R_{23}(v)R_{13}(u+v)R_{12}(u)\label{YB}
\end{equation}
where the arguments u and v, called spectral parameters, have values
in C. The solution of the YBE is an R matrix defined in End (V \ensuremath{\otimes}
V ). The indexed matrices $R_{ij}$ appearing in (1) are defined in
End (V \ensuremath{\otimes} V \ensuremath{\otimes} V ) through the
formulas 
\begin{equation}
R_{12}=R\otimes I,\qquad R_{23}=I\otimes R,\qquad R_{13}=P_{23}R_{12}P_{23}
\end{equation}
where $I\in End(V$) is the identity matrix, $P\in End(V\otimes V)$
is the permutation matrix (defined by the relation $P(A\otimes B)P=B\otimes A$
for$ $ $A,B\in End(V))$ and $P_{12}=P\otimes I,P_{23}=I\otimes P$.

From the quantum group invariant representation for non-exceptional
affine Lie algebra $A_{n-1}^{(1)}$ \cite{jimbo},we consider the
following generalization for a R-matrix solution

\begin{equation}
R(u)=\sum_{i=1}^{n+1}a_{ii}(u)\mathrm{e}_{ii}\otimes\mathrm{e}_{ii}+\sum_{i\neq j}^{n+1}b_{ij}(u)\mathrm{e}_{ii}\otimes\mathrm{e}_{jj}+\sum_{i\neq j}^{n+1}c_{ij}(u)\mathrm{\mathrm{e}}_{ij}\otimes\mathrm{e}_{ji}\label{eq:R-1}
\end{equation}
where $e_{ij}$ are the Weyl matrices $(e_{ij})_{ab}=\delta_{ia}\delta_{jb},$
acting for a $n+1$ dimensional vector space $V$ at the site $n$
. The R-matrix elements $a_{ii}(u)$, $b_{ij}(u)$ and $c_{ij}(u)$
are fixed by the YBE. Breaking of the original symmetry releases several
parameters.

At this point we observe that (\ref{eq:R-1}) no longer has $A_{n-1}^{(1)}$
symmetry. But using the fact that there are many free parameters the
particular choice
\begin{equation}
a_{ij}(u)=a(u),\quad b_{ij}(u)=b(u),\;c_{ij}(u)=\mathbf{c}\qquad(i<j),\quad c_{ij}(u)=\mathbf{c}\exp(u)\label{an}
\end{equation}
where $\mathbf{c}$ is a constant, we are recovering the $A_{n-1}^{(1)}$
symmetry\cite{jimbo}. Thus, we understand that the result presented
in this work with the choice (\ref{an}) reproduces all those solved
with the $A_{n-1}^{(1)}$ symmetry.

For each value of $n$>1, the R matrix has $n+1$ diagonal entries
$a_{ii}(u)$, that are determined by recurrence relative to $a_{11}(u)$

\begin{equation}
a_{kk}(u)=a_{11}(u)+(\alpha_{kk}-\alpha_{11})F(u),\qquad k=2,...,n+1
\end{equation}
where 
\begin{equation}
\alpha_{jj}=\left(\frac{d}{du}a_{jj}(u)\right)_{u=0}\quad and\quad F(u)=\frac{e^{\kappa_{1}u}-e^{\kappa_{2}u}}{-\kappa_{2}+\kappa_{1}}\label{F}
\end{equation}
where 
\begin{equation}
\kappa_{1}=\mu_{13}+\mu_{21}-\mu_{23}
\end{equation}
and 
\begin{equation}
\kappa_{2}=-\mu_{13}+\mu_{12}+\mu_{23}.
\end{equation}

The remaining $n(n+1)$diagonal entries 
\begin{equation}
b_{ij}(u)=\beta_{ij}F(u),\qquad i\neq j=\{1,...,n+1\}
\end{equation}
where 
\begin{equation}
\beta_{ij}=\left(\frac{d}{du}b_{ij}(u)\right)_{u=0}
\end{equation}
with the constraints 
\begin{equation}
\beta_{ji}\beta_{ij}=\beta_{21}\beta_{12}=(\kappa_{1}-\alpha_{11})(\kappa_{2}-\alpha_{11})
\end{equation}
The number of fixed $\beta_{ij}$ is $\frac{1}{2}$ $n-1)(n+2)+1$.

The $n(n+1)$ off-diagonal matrix elements 
\begin{equation}
c_{ij}(u)=e^{\mu_{ij}u}\label{C}
\end{equation}
All $\mu_{ij}$ of the $c_{ij}(u)$ below od main diagonal are fixed
by the relation 
\begin{equation}
\mu_{ji}=\kappa_{1}+\kappa_{2}-\mu_{ij},\qquad j>i
\end{equation}
The number of fixed $\mu_{ji}$ is$\frac{1}{2}$ $(n-1)(n+2)$and
some $\mu_{ij}$ above the main diagonal can be fixed by two relation
\begin{equation}
\mu_{ij}=\kappa_{1}-\mu_{1i}+\mu_{1j}\mathrm{\quad and}\quad\mu_{ij}=\kappa_{2}-\mu_{1i}+\mu_{1j},\quad j<i\label{mu}
\end{equation}
The number of fixed parameters $\mu_{ij}$ is $(n+1)(n-2)/2$. It
means that we have two sets of solutions. With these relations the
Yang-Baxter equation and its derivatives are solved by two sets of
$2^{n}$ solutions of the following $n$ equations 
\begin{equation}
(\alpha_{kk}-\alpha_{11})(\alpha_{kk}+\alpha_{11}-\kappa_{1}-\kappa_{2})=0
\end{equation}

Therefore we have two differentes values for the $a_{kk}(u)$ and
$n$ parameters $\alpha_{ii}$ are fixed. It mean that our R matrix
solutions have $n(n+3)/2+3$ free parameters

\begin{equation}
a_{kk}(u)=A(u)=\frac{(\kappa_{1}-\alpha_{11})\exp(\kappa_{2}u)-(\kappa_{2}-\alpha_{11})\exp(\kappa_{1}u)}{\kappa_{1}-\kappa_{2}}\label{A}
\end{equation}
For $\alpha_{ii}=\alpha_{11}$ and 
\begin{equation}
a_{kk}(u)=B(u)=\frac{(\kappa_{1}-\alpha_{11})\exp(\kappa_{1}u)-(\kappa_{2}-\alpha_{11})\exp(\kappa_{2}u)}{\kappa_{1}-\kappa_{2}}
\end{equation}
For $\alpha_{kk}=\kappa_{1}+\kappa_{2}-\alpha_{11}$.

As presented and discussed in \cite{vieira0}, the $2^{n}$ solutions
have been reduced for $n+1$ different $R$ matrices

\section{The Reflection Equations}

In the current literature we can find several works on the solutions
of the reflection equations with $A_{n-1}$symmetry, this is with
(\ref{an}). Among them we cite \cite{de Vega,Abad,ganden,lima2,Yang,vidas}

While the YBE reading of the conservation of the linear moment , the
RE reading of the conservation of energy.

Given a solution R of YBE, we can build a monodromy matrix $T(u)$,
whose trace defines a row to row transfer matrix $\tau(u).$

The existence of a K matrix allows us to reflect T(u) on a wall. The
reflected $T(-u)^{-1}$ which in turn will be reflected on a second
wall by the matrix $K^{+}$, dual of K. Thus we have the Sklyanin's
transfer matrix \cite{sklyanin}

\begin{equation}
t(u)=Tr_{a}(K^{^{+}}(u)T(u)K(u)T^{-1}(-u))
\end{equation}
where $Tr_{a}$means trce in the auxiliary space.

In our case there is an isomorphism \cite{nepo}

\begin{equation}
K^{+}(u)=K(-u-\rho)^{t}M
\end{equation}
where $\rho$ is the crossing parameter,$M$ is the crossing matrix
and $t$ means tranposition.

The translational invariance in the vertical (temporal) direction
of the monodromy on the wall with $K(u)$ is given by the RE \cite{sklyanin,nepo}:

\begin{equation}
RE=R(u-v)K_{1}(u)PR(u+v)PK_{2}(v)=K_{2}(v)-K_{2}(v)R(u+v)K_{1}(u)PR(u-v)P
\end{equation}
where $K_{1}=K\otimes I$ , $K_{2}=I\otimes K$ and $P=R(0)$ .

This functional equation can be properly transformed into an algebraic
equation, through its derivatives with respect to the parameter v
at point $v=0$.

\begin{equation}
DRE=-B_{2}R(u)K_{1}(u)PR(u)P+R(u)K_{1}(u)PR(u)PB_{2}-2D(u)K_{1}(u)PR(u)P+2R(u)K_{1}(u)PR(u)P\label{dre}
\end{equation}
Similarly we have another equation for the derivatives with respect
to $u$, but unlike YBE, we just need to use (\ref{dre})

The diagonal matrices have the form.

To obtain (18) we have used the definitions

\begin{equation}
D(u)=\left[\frac{\partial R(u\pm v)}{\partial v}\right]_{v=0},\qquad\qquad B=\left[\frac{dK(v)}{dv}\right]_{v=0}
\end{equation}
and the regular condition $K(0)=I$.

From (\ref{eq:R-1}) we get the derivative of $R(u)$:

\begin{equation}
D(u)=\sum_{i=1}^{n+1}a_{ii}(u)'\mathrm{e}_{ii}\otimes\mathrm{e}_{ii}+\sum_{i\neq j}^{n+1}b_{ij}(u)'\mathrm{e}_{ii}\otimes\mathrm{e}_{jj}+\sum_{i\neq j}^{n+1}c_{ij}(u)'\mathrm{\mathrm{e}}_{ij}\otimes\mathrm{e}_{ji}
\end{equation}
For a given $n$ the K-matrix is given by and its derivative at the
point $v=0$ are 
\begin{equation}
K(u)=\sum_{i=1}^{n+1}k_{ii}(u)\,\mathrm{e}(i,i)\qquad B=\sum_{i=1}^{n+1}\theta_{ii}(u)\,\mathrm{e}(i,i)\label{KD}
\end{equation}
where $\mathrm{e}(i,j)$ are $n+1$ by $n+1$ matrices with entries
$[e(i,j)]_{ab}=\delta_{ai}\delta_{bj}$.

Taking into account the differential equation (\ref{dre}) with (\ref{KD})
and the date of the n+1 solutions of the correponding Yang-Baxter
equations discussed in \cite{vieira0}.

\subsection{Case n=2}

In this case 
\begin{equation}
K(u)=\left(\begin{array}{ccc}
k_{11}(u) & 0 & 0\\
0 & k_{22}(u) & 0\\
0 & 0 & k_{33}(u)
\end{array}\right)\qquad B=\left(\begin{array}{ccc}
\theta_{_{11}} & 0 & 0\\
0 & \theta_{22} & 0\\
0 & 0 & \theta_{33}
\end{array}\right)
\end{equation}
This completes the data for (18)

From the equation $\textrm{DRE}[$2,4{]} we find $k_{22}(u)$

\begin{equation}
k_{22}(u)=\frac{[(2\mu_{21}+\theta_{11}-\theta_{22})F(u)-2F'(u)]c_{21}(u)k_{11}(u)}{[(2\mu_{12}+\theta_{22}-\theta_{11})F(u)-2F'(u)]c_{12}(u)}
\end{equation}

From the equation $\textrm{DRE}[3,7]$we get $k_{33}(u)$

\begin{equation}
k_{33}(u)=\frac{[(2\mu_{31}+\theta_{11}-\theta_{33})F(u)-2F'(u)]c_{31}(u)k_{11}(u)}{[(2\mu_{13}+\theta_{33}-\theta_{11})F(u)-2F'(u)]c_{13}(u)}
\end{equation}
where $c_{ij}$, $F$ and $F'$ are given by (\ref{C}) and (\ref{F})

Now, from $\textrm{DRE}[6,8]$ and $\textrm{DRE}[8,6]$ we find two
constraints 
\begin{equation}
eq_{1}=(-\theta_{33}+\theta_{11}+2\mu_{21}-2\mu_{23})(-\theta_{33}+\theta_{22}+2\mu_{12}-2\mu_{13})
\end{equation}
\begin{equation}
eq_{2}=(-\theta_{11}+\theta_{22}+2\mu_{13}-2\mu_{23})(-\theta_{33}+\theta_{11}+2\mu_{21}-2\mu_{23})
\end{equation}
which provides two solutions:

\begin{align}
\theta_{22}= & \theta_{11}-2\mu_{13}+2\mu_{23}\\*
\nonumber \\
\theta_{33}= & \theta_{11}-2\mu_{23}+2\mu_{21}\nonumber \\
\theta_{33}= & \theta_{22}+2\mu_{12}-2\mu_{13}
\end{align}

Note that in the solutions we have one value for $\theta_{22}$ and
two values for $\theta_{33}.$

Being more explicit 
\begin{equation}
K_{1}=\left(\begin{array}{ccc}
k_{11} & 0 & 0\\
0 & k_{11}\mathrm{e}^{^{(-2u(\mu_{13}-\mu_{23})}} & 0\\
0 & 0 & k_{11}\mathrm{e}^{(2u(\mu_{21}-\mu_{23})}
\end{array}\right)
\end{equation}
and 
\begin{equation}
K_{2}=\left(\begin{array}{ccc}
k_{11} & 0 & 0\\
0 & k_{11}\mathrm{e}^{(-2u(\mu_{13}-\mu_{23})} & 0\\
0 & 0 & k_{11}\mathrm{e^{(2u(\mu_{12}-2\mu_{13}+\mu_{23})}}
\end{array}\right)
\end{equation}
where $k_{11}$=$k_{11}(u)$ is an arbitrary function with $k_{11}(0)=1$
and $k'_{11}(0)=\theta_{11}.$

With an appropriate choice of free parameters, for example, if we
fix $\mu_{23}$ as $\mu_{23}=\mu_{13}+\frac{1}{2}(\mu_{_{21}}-\mu_{12})$we
will get $K_{2}=K_{1}$, giving us a better understanding of the free
parameters.

\subsection{Case n=3 }

Now we have

\begin{equation}
K(u)=\left(\begin{array}{cccc}
k_{11}(u)\\
 & k_{22}(u)\\
 &  & k_{33}(u)\\
 &  &  & k_{44}(u)
\end{array}\right),\quad B=\left(\begin{array}{cccc}
\theta_{11}\\
 & \theta_{22}\\
 &  & \theta_{33}\\
 &  &  & \theta_{44}
\end{array}\right)
\end{equation}

From $\text{\ensuremath{DRE[2,5]}},$$\text{\ensuremath{DRE[3,9]}}$
and $DRE[4,13]$ we obtain $k_{22}$, $k_{33}$ and $k_{44}$, respectively

\begin{equation}
k_{22}(u)=\frac{[(2\mu_{21}+\theta_{11}-\theta_{22})F(u)-2F'(u)]c_{21}(u)k_{11}(u)}{[(2\mu_{12}+\theta_{22}-\theta_{11})F(u)-2F'(u)]c_{12}(u)},
\end{equation}
\begin{equation}
k_{33}(u)=\frac{[(2\mu_{31}+\theta_{11}-\theta_{33})F(u)-2F'(u)]c_{31}(u)k_{11}(u)}{[(2\mu_{13}+\theta_{33}-\theta_{11})F(u)-2F'(u)]c_{13}(u)}.
\end{equation}
and 
\begin{equation}
k_{44}(u)=\frac{[(2\mu_{41}+\theta_{11}-\theta_{44})F(u)-2F'(u)]c_{41}(u)k_{11}(u)}{[(2\mu_{14}+\theta_{44}-\theta_{11})F(u)-2F'(u)]c_{14}(u)}
\end{equation}

Using these amplitudes the remaining equations,$\textrm{DRE}[7,10],$$\textrm{DRE}[8,14]$
and $\textrm{DRE}[12,15]$, leave us with three constraints , respectively
\[
eq_{1}=(2\mu_{21}-2\mu_{23}+\theta_{11}-\theta_{33})(2\mu_{13}-2\mu_{23}-\theta_{11}+\theta_{22})(2\mu_{12}-2\mu_{13}+\theta_{22}-\theta_{33})
\]
\[
eq_{2}=(2\mu_{13}-2\mu_{23}-\theta_{11}+\theta_{22})(2\mu_{13}-2\mu_{14}-\theta_{44}+\theta_{22})(2\mu_{13}-2\mu_{23}+2\mu_{21}-2\mu_{14}+\theta_{11})
\]
\begin{equation}
eq_{3}=(2\mu_{14}-2\mu_{13}+\theta_{44}-\theta_{33})(2\mu_{12}-4\mu_{13}+2\mu_{23}+\theta_{11}-\theta_{33})(2\mu_{14}-2\mu_{13}+2\mu_{23}-2\mu_{21}+\theta_{44}-\theta_{11})\label{eqs}
\end{equation}

For the solutions of this set of equation, we have find one $\theta_{22},$
two $\theta_{33}$and three $\theta_{44}$ , therefore, six solutions:

\begin{align}
\theta_{22}= & \theta_{11}-2\mu_{13}+2\mu_{23}\\*
\nonumber \\
\theta_{33}= & \theta_{11}+2\mu_{21}-2\mu_{23}\nonumber \\
\theta_{33}= & \theta_{22}+2\mu_{12}-2\mu_{13}\\*
 & *\nonumber \\
\theta_{44}= & \theta_{11}+2\mu_{13}+2\mu_{21}-2\mu_{23}-2\mu_{14}\nonumber \\
\theta_{44}= & \theta_{22}+2\mu_{12}-2\mu_{14}\nonumber \\
\theta_{44}= & \theta_{33}+2\mu_{13}-2\mu_{14}
\end{align}

Now let's look at the parameters $\theta_{ii}$ for the case $n=3$
are the same as the case $n=2$ plus more three new parameters $\theta_{44}$
. It means that each of the $3$ by $3$ matrices of the case $n=2$
are contained in three 4 by 4 matrices for the case $n=3$. Due to
the free parameters, coming from the $R$ matrix.

The new three values for $k_{44}(u)$ are 
\begin{alignat}{1}
k_{44_{1}}(u)= & e^{2u(\mu_{13}-\mu_{14}+\mu_{21}-\mu_{23})}k_{11}(u)\nonumber \\
k_{44_{2}}(u)= & e^{2u(\mu_{12}-\mu_{14}-\mu_{13}-+\mu_{23})}k_{11}(u)\nonumber \\
k_{44_{3}}(u)= & \mathrm{e^{2u(\mu_{13}-\mu_{14}+\mu_{21}-\mu_{23})}k_{11}(u)}
\end{alignat}
for $K_{1}$ and for $K_{2}$.

Note that while we know that $k_{44_{1}}=k_{44_{3}}$ we will take
this into account only later

This construction will now be extended to the other vertex models.

\subsection{Case $n>3$}

Now we will present a generalization for these results for a given
value of $n$

Since there are two sets of $n+1$ matrices R different from each
other by the values of $\mu_{ij}$ with $i<j$ (\ref{mu}). Let's
choose the set with $a_{11}(u)=A(u)$ \ref{A} and $\mu_{ij}=\kappa_{2}-\mu_{1i}+\mu_{ij}$.

From the $(n+1)^{4}$entries of (18) we will only need $n+1$ in order
to find the $k_{ii}$matrix entries, namely

\begin{equation}
\mathrm{DRE}[a,(i-1)(n+1)+1],\Longrightarrow k_{ii}(u),\qquad i=2,3,...,n+1
\end{equation}

After that, we can use the function 
\begin{equation}
V(i,j,u)=((2\mu_{ij}-\theta_{jj}+\theta_{ii})F(u)-2F'(u))c_{ij}(u)\label{os V}
\end{equation}
to write all the diagonal elements of the reflection matrices for
all values of $n$ in a general form 
\begin{equation}
k_{ii}(u)=\frac{V(i,1,u)}{V(1,i,u)}k_{11}(u),\qquad\qquad a=2,3,...,n+1
\end{equation}
where $k_{11}(u)$ is an arbitrary function such that $k_{11}(0)=1$
and $\left[\frac{d}{du}k_{11}(u)\right]_{u=0}=\theta_{11}$.

From these data we see that they do not depend on $a_{11}(u)$ and
therefore are also valid for $a_{11}(u)=B(u)$ . It means the solutions
for the $n+1$ matrices of the set considered. Moreover, everything
is valid for the second set as long as $\kappa_{2}$ is exchanged
for $\kappa_{1}$or taking $\mu_{23}=\mu_{13}+\frac{1}{2}(\mu_{21}-\mu_{12})$.
Considerably increasing the number of solutions.

Thus, and we are left with the task to find the parameters $\theta_{\alpha\alpha}$
depend on $\theta_{11}$ and $\mu_{ij}$ the free parameters of the
$n+1$ $R$-matrices, solutions of YBE.

This task takes a lot of time,but we have the reward of being able
to get all solutions with the help of the following expression 
\begin{equation}
\theta_{ii}=f_{ijk}=\theta_{jj}+2\mu_{ik}-2\mu_{jk}\qquad\qquad(i>j)\label{sol}
\end{equation}
The index $k$ is special, it has to be different from $i$ and $j$,
it means that there are $n-1$ possible values to be assigned to $k$.

It is clear that for high values of $n$ it is difficult to determine
the value of $k$. We managed to get rid of this difficulty:

As already described for the first two values of n, for a given value
of $n$, we only need to calculate the $n$ $\theta_{n+1,n+1}$because
all others$(n-1)!$ $\theta_{ii}$ have already been calculated. The
inclusion of these $n$ new values, defines the $n!$ diagonal solutions
of the our vertex model. This is done by choosing the smallest among
the possible values of $k$.

Thus the two solutions for $n=2$ are

\begin{equation}
\begin{array}{cccc}
1: & \theta_{22}=f_{213} & , & \theta_{33}=f_{312}\\
2: & \theta_{22}=f_{213} & , & \theta_{33}=f_{321}
\end{array}
\end{equation}
and the six solution for $n=3$ are

\begin{equation}
\begin{array}{ccccccc}
1: &  & \theta_{22}=f_{213} &  & \theta_{33}=f_{312} &  & \theta_{44}=f_{412}\\
2: &  & \theta_{22}=f_{213} &  & \theta_{33}=f_{312} &  & \theta_{44}=f_{421}\\
3: &  & \theta_{22}=f_{213} &  & \theta_{33}=f_{312} &  & \theta_{44}=f_{431}\\
\\
4: &  & \theta_{22}=f_{213} &  & \theta_{33}=f_{321} &  & \theta_{44}=f_{412}\\
5: &  & \theta_{22}=f_{213} &  & \theta_{33}=f_{321} &  & \theta_{44}=f_{421}\\
\mathbf{6}: &  & \theta_{22}=\mathbf{f}_{213} &  & \theta_{33}=\mathbf{f}_{321} &  & \theta_{44}=\mathbf{f}_{431}
\end{array}
\end{equation}
Bold identifies the repeated solution.So, we found five different
solutions from each other

To better understand this approach, we will also present the solutions
for case $n=4$

The $n+1$ diagonal entries of the $K$ -matrix 
\begin{equation}
\begin{array}{ccc}
k_{11}(u) & = & 1\,k_{11}(u)\\
k_{22}(u) & = & \frac{V(2,1,u)}{V(1,2,u)}\,k_{11}(u)\\
k_{33}(u) & = & \frac{V(3,1,u)}{V(1,3,u)}\,k_{11}(u)\\
k_{44}(u) & = & \frac{V(4,1,u)}{V(1,4,u)}\,k_{11}(u)\\
k_{55}(u) & = & \frac{V(5,1,u)}{V(1,5,u)}\,k_{11}(u)
\end{array}\label{K5}
\end{equation}
where the $V(i,1,u)$,$\qquad i=2,3,4,5$ are given by (\ref{os V})

The next step is to find the parameters $\theta_{ii}$ contained in
\ref{K5}, that are already given by (\ref{sol}).

After finding the $\theta_{ii}$ we have to identify which and how
many are repeated. In terms of $\mathbf{f}_{ijk}$ they are those
with $i=n+1$, $j=3,4,..,n$ and $k$ obeying the rule the smallest
possible values.

For a given $n$ we have 
\begin{equation}
S(n)=\frac{1}{2}(n-2)(n-1)
\end{equation}
It means we've found 
\begin{equation}
T(n)=n!-S(n)
\end{equation}
different solutions.

Agreeing with what we have previously presented: $T(2)=2,\qquad T(3)=5$.
It is at least interesting to explicitly find $T(4)=$ $21$ solutions
for the case $n=4$ case.

\section{Conclusion}

One of the advantages of diagonal solutions is that the spectrum of
their Hamiltonian with diagonal boundaries are directly solved by
the Bethe ansätze \cite{doikou}, see also \cite{fire}. 

The matrices $K$ diagonal solutions of (RE), presented in this work
have a peculiar behavior: For a given value of $n$ its $n+1$ by
$n+1$ diagonal is the $n$ by $n$ digonal of $n-1$ case plus the
element $k_{n+1\,n+1}$$(u)$ . Thus, as the approach used is valid
for $n=1$, $k_{11}(u)$ and $k_{22}(u)$ define the diagonal solution
of the six vertex- model.

This is also expected for non-diagonal solutions, which we hope to
presented in the future. The calculations of non-diagonal $K$ matrices
are more extensive and require a lot of time and their generalization
is not as simple as in the diagonal case, see for example the case
$n=2$ as presented in \cite{vieira1}

\appendix

\section{An essay on non-diagonal solutions}

Going back to the reflection equation (\ref{dre}) but now with

\begin{equation}
K(u)=\sum_{i,j=1}^{n+1}k_{ij}(u)\,\mathrm{e}(i,j)\qquad,K(0)=I,\quad B=\sum_{i,j=1}^{n+1}\theta_{ij}(u)\,\mathrm{e}(i,j)
\end{equation}
The equations $DRE[i,i]=0$ are all resolved by relations
\begin{equation}
\theta_{ij}k_{ji}(u)=\theta_{ji}k_{ij}(u),\quad(i\neq j)\label{sym}
\end{equation}

Now we choose a particular $k_{ij}(u)$ to be different from zero,
wth $\theta_{ij}\neq0$ and write all remaining elements in terms
of the $k_{ij}(u)$ and from the remaining equations we will find
a very strong restriction for the matrix elements: for a particular
$k_{ij}(u)\neq0$ the elements different from zero are $k_{ii}(u),k_{jj}(u)$
and $k_{ji}(u)$ .

\begin{equation}
S=\{k_{ij}(u)\neq0,\quad k_{ji}(u)\neq0,\quad k_{ii}(u)\neq0,\quad k_{jj}(u)\neq0\}\label{vinforte}
\end{equation}
these four elements different from zero define a block within the
matrix $K(u)$

\begin{equation}
B_{ab}=\left[\begin{array}{ccc}
k_{aa} & \ldots & \mathbf{k}_{ab}\\
\vdots & \ddots & \vdots\\
k_{ba} & \ldots & k_{bb}
\end{array}\right]
\end{equation}
 where $a<b$ . The number of blocks $B_{ab}$ within K define the
types of solutions

We note that all entries of are diagonal and also non-zero due to
the normal condition.

We identify the$K(u)$ solution with one block $B_{ab}$ by $K_{ab}^{[I]}$(u)
and with two blocks by $K_{ab}^{[II]}(u)$ 

The solutions with one block have the form

\begin{equation}
K_{ab}^{[I]}(u)=k_{ab}(u)\mathrm{e}(a,b)+k_{ba}\mathrm{e}(b,a)+\sum_{i}^{n+1}k_{ii}\mathrm{e}(i,i),\qquad,a<b,\{(a=1,..,n),\qquad(b=2,...,n+1)\}
\end{equation}
we count $\frac{n}{2}(n+1)$ $K_{ab}^{[I]}$ matrices each with $n+3$
non-null elements.

For six-vertex model we have one solution
\begin{equation}
K_{12}^{[I]}=\left(\begin{array}{cc}
k_{11} & \mathbf{k}_{12}\\
k_{21} & k_{22}
\end{array}\right),
\end{equation}
for fifteen-vertex model we have three solutions
\begin{equation}
K_{12}^{[I]}=\left(\begin{array}{ccc}
k_{11} & \mathbf{k}_{12} & 0\\
k_{21} & k_{22} & 0\\
0 & 0 & k_{33}
\end{array}\right),\;K_{13}^{[I]}=\left(\begin{array}{ccc}
k_{11} & 0 & \mathbf{k}_{13}\\
0 & k_{22} & 0\\
k_{31} & 0 & k_{33}
\end{array}\right),\;K_{23}^{[I]}=\left(\begin{array}{ccc}
k_{11} & 0 & 0\\
k_{21} & k_{22} & \mathbf{k}_{23}\\
0 & k_{32} & k_{33}
\end{array}\right)
\end{equation}

We also found a second type of solution containing two blocks $B_{ab}$
and $B_{cd}$ defining a second type of solution $K_{ab}^{[II]}$.
The existence of blocks implies the following condition for non-diagonal
entries
\begin{equation}
k_{ij}(u)k_{ji}(u)=k_{rs}(u)k_{rs}(u)
\end{equation}
since 
\begin{equation}
i+j=r+s,\mathrm{\qquad mod\:}(n+1)\label{mod}
\end{equation}
As $i<j$ and $r<s$ the choice $r=i-1,s=j+1$ is enough to solve
(\ref{mod}) and we find the Following form for $K_{ab}^{[II]}$ :
\[
K_{ab}^{]II]}(u)=k_{ab}(u)\mathrm{\,e}(a,b)+k_{ba}(u)\mathrm{\,e}(b,a)+k_{a+b,n+1}(u)\mathrm{e}(a+b,n+1)+k_{n+1,a+b}(u)\mathrm{\,e}(n+1,a+b)
\]
\begin{equation}
+\sum_{i=1}^{n+1}k_{ii}(u)\,\mathrm{e}(i,i)
\end{equation}
for $a+b\leq n$. 

For $a+b\geq n+1$ the solution has the form
\[
K_{ab}^{]II]}(u)=k_{ab}(u)\mathrm{\,e}(a,b)+k_{ba}(u)\mathrm{\,e}(b,a)+k_{a+1,b-1}(u)\mathrm{e}(a+1,b-1)+k_{b-1,a+1}(u)\mathrm{\,e}(b-1,a+1)
\]
\begin{equation}
+\sum_{i=1}^{n+1}k_{tt}(u)\,\mathrm{e}(i,i)
\end{equation}

The $28$-vertex model has six type-I solutions and two type-II solutions
\begin{equation}
K_{12}^{[II]}u)=\left(\begin{array}{cccc}
k_{11} & \mathbf{k}_{12} & 0 & 0\\
k_{21} & k_{22} & 0 & 0\\
0 & 0 & k_{33} & \mathbf{k}_{34}\\
0 & 0 & k_{43} & k_{44}
\end{array}\right),K_{14}^{[II]}=\left(\begin{array}{cccc}
k_{11} & 0 & 0 & \mathbf{k}_{14}\\
0 & k_{22} & \mathbf{k}_{23} & 0\\
0 & k_{32} & k_{33} & 0\\
k_{41} & 0 & 0 & k_{44}
\end{array}\right)
\end{equation}
We found $\frac{1}{2}(n+1)(n-2)$ of these type-II solutions.

In reference \cite{Yang} the non-diagonal trigonometric solutions
for the $A_{n-1}^{(1)}$ models were classified and in reference \cite{vidas}
the rational case is addressed. 

We will present the non-diagonal solutions elsewhere.

Let's conclude this essay with some additional information:
\begin{itemize}
\item Our $n+1$ $R$ matrices have the same shape as the $R$ matrix of
algebra $A_{n-1}^{(1)}$, hence our $\text{K}$ matrices, but with
different entries.
\item In the calculations of the $\{K_{ab}^{(I]}(u),K_{ab}^{(II]}(u)\}$
we need to bind some free parameters of $R$.
\item For a given value of $n$, the solution $K_{ab}^{[I]}(u)$ contains
the $K_{ab}^{[I]}(u)$ of the previous values of $n$. In the limit
$n=1$ , $K_{ab}^{[I]}(u)$ is the solution for the six-vertex. Its
well-known solution is reproduced in the reduction (\ref{an}). 
\item Even with so many free parameters it was not possible to find a solution
with all non-zero entries.
\item We still have nothing to write about the possibilities of finding
solutions type -III, as well as the possibilities to consider other
types of blocks
\end{itemize}

\end{document}